\documentclass[12pt]{article}
\usepackage{amsmath,epsfig,graphicx,amssymb}
\usepackage[margin=0.7in]{geometry}
\newcommand{\beq}{\begin{equation}}
\newcommand{\eeq}{\end{equation}}
\newcommand{\ben}{\begin{eqnarray}}
\newcommand{\een}{\end{eqnarray}}
\newcommand{\bes}{\begin{subequations}}
\newcommand{\ees}{\end{subequations}}
\newcommand{\bFig}{\begin{figure}}
\newcommand{\eFig}{\end{figure}}
\usepackage{hyperref}
\date{}
\begin{document}
\title{Continuous Transitions Between Quantum and Classical Motions}
\author{Partha Ghose\footnote{partha.ghose@gmail.com} \\
The National Academy of Sciences, India,\\ 5 Lajpatrai Road, Allahabad 211002, India,\\and\\
Klaus von Bloh, Lilienthal, Germany \footnote{kvbloh@t-online.de}.}
\maketitle
\begin{abstract}
Using a nonlinear Schr\"{o}dinger equation for the wave function of all systems, continuous transitions between quantum and classical motions are demonstrated for (i) the double-slit set up, (ii) the 2D harmonic oscillator and (iii) the hydrogen-like atom, all of which are of empirical interest.
\end{abstract}
\section{Introduction}
The problem of continuity of mathematical and conceptual descriptions of classical and quantum systems was discussed and a possible solution was offered by one of the authors in 2002 in a couple of papers \cite{ghose1,ghose2}. It was pointed out that the causal and ontological interpretation of quantum mechanics \cite{bohm,Holland} provided a suitable basis to view smooth transitions between these systems particularly clearly. Since then interest has grown in the area of mesoscopic systems in chemistry and solid state physics which exhibit both quantum and classical features, and the solution suggested in \cite{ghose1} and \cite{ghose2} acquires relevance. Here we use the word `mesoscopic' to mean something that is neither fully classical nor fully quantum mechanical. Hence, a macroscopic system can be mesoscopic in this sense. 

The purpose of this paper is two-fold: (i) to summarise the key points of the argument and suggest a simpler mathematical framework that can be used for both relativistic and non-relativistic systems, and (ii) give three additional examples of smooth transitions of direct empirical relevance. The first one is the demonstration of the famous double-slit experiment where the transition from classical trajectories to Bohmiam trajectories giving rise to an interference pattern as the coupling of the system to the environment is reduced is clearly visible. The second example is that of a 2D harmonic oscillator which has numerous possibilities of applications in physics and chemistry. The third example is that of some electron states in the hydrogen atom, their Bohmian trajectories and how they go over smoothly to classical Keplerian orbits. It is hoped that these examples will stimulate further and more detailed investigations of other mesoscopic systems using the same technique.

It is well known that the conceptual boundary between classical and quantum descriptions of a physical system is not only arbitrary, there is an obvious disconnect between these two descriptions. In the ultimate analysis the classicality of the measuring apparatus is required to extract physically observable predictions from Schr\"{o}dinger evolution of the wave function, but this classicality cannot be deduced from an underlying quantum mechanical substratum. Attempts have been made to show how the classical world emerges from an underlying quantum mechanical susbstratum through decoherence \cite{joos,Zurek1,Zurek2}, but these approaches, though practically very useful, work only in the pointer basis---quantum coherence persists in other equivalent bases, and the classicality achieved is limited. It does not really bridge the conceptual gap between truly classical and quantum systems. 

The classical description is both realistic and causal whereas the standard quantum mechanical description is neither realistic nor causal. The de Broglie-Bohm interpretation of quantum mechanics showed that empirical evidence does not compel one to renounce realism and causality. However, being only an interpretation of quantum mechanics, its logic was not compelling, and by and large the community of physicists dismissed it as a curious hangover from the classical age. Furthermore, there was no clear way even in this interpretation to see the link between classical and quantum descriptions except in so far as it enables a complete isolation of all quantum effects in the quantum potential. There is nothing within the theory to suggest when and how the quantum potential gets switched off as a system approaches the classical limit. In fact, the nonlocality and contexuality of the interpretation (most strongly reflected in the quantum potential \cite{BH}) made a smooth transition look even harder to achieve than in the standard interpretation.

The main point of \cite{ghose1} and \cite{ghose2} was that all quantum characteristics of a system get completely quenched when it is strongly coupled to its environment. As the coupling with the environment is reduced, the quantum characteristics of the system start to emerge, and finally the system becomes fully quantum mechanical when completely isolated from its environment. This simple idea can be implemented by introducing the concept of a wave function for all systems, including those that are classical, which obeys a certain nonlinear equation. 

\section{Dynamics of Mesoscopic Systems}
It was Koopmann \cite{koop} and von Neumann \cite{vN} who first introduced the idea of complex wave functions for classical systems that are square integrable and span a Hilbert space. Though the classical wave functions $\phi(q,p)$ are complex, their relative phases are unobservable in classical mechanics. This is achieved by postulating that the wavefunctions obey Liouville equations, and showing that the density function in phase space $\rho(q,p) = \phi^*(q,p)\phi(q,p)$ also obeys the Loiuville equation. Thus, the classical phase space probability density and its dynamics can be recovered from the dynamics of the underlying complex wavefunctions $\phi$ and $\phi^*$. The dynamical variables $p$ and $q$ are assumed to be commuting operators. The method can be generalized to the case of the electromagnetic field \cite{ghose3}

While we use the idea of square integrable complex wave functions for all systems, we postulate a different dynamics, namely a modified Schr\"{o}dinger equation of the form
\ben 
i\,\hbar\,\frac{\partial \psi}{\partial t} &=&
\left(-\frac{\hbar^2}{2 m}\, \nabla^2  + V(x)  - \lambda(t)Q \,\right)\,\psi \label{1},\\
Q &=& - \frac{\hbar^2}{2 m}\,\frac{\nabla^2 R}{R},\label{2}
\een 
where $R$ is defined by writing $\psi$ in the polar form $\psi = R{\rm exp}\left(\frac{i}{\hbar}S \right)$. The term $\lambda(t)Q$ represents the nonlinear coupling of the system to its environment. Separating the real and imaginary parts, one gets the modified Hamilton-Jacobi equation
\beq
\partial S/\partial t + \frac{(\nabla S)^2}{2m} + V + (1 - \lambda(t))Q= 0,\label{3}
\eeq
for the phase $S$ of the wave function, and the continuity equation

\beq
\frac{\partial \rho ( {\bf x}, t )}{\partial t} +
{\bf \nabla}\,.\, (\,\rho\, \frac{{\bf \nabla}\,S}{m}) = 0 \label{4}
\eeq
for the density $\rho = \psi^*\psi = R^2$. The coupling function $\lambda(t)$ is chosen such that the system behaves fully classically in the limit $\lambda(t)\rightarrow 1$ and fully quantum mechanically in the limit $\lambda(t)\rightarrow 0$. Putting ${\bf p} = {\bf \nabla}\, S$, one can see that the functions $\rho = R^2$ and $S$ obeying the differential equations (\ref{3}) and (\ref{4}) are completely decoupled in the classical limit, ensuring no interference effects. Note that Planck's constant $h$ drops out from these equations in the classical limit. 

Since the momentum is defined by $p_i = \partial_i S$, one has
\beq
u_i = \frac{d x_i}{dt} = \frac{1}{m}\partial_iS,\label{5} 
\eeq
and integration of this first-order equation gives the set of trajectories corresonding to an arbitrary choice of initial positions. In Bohmian mechanics, the initial positions are chosen to fit the quantum mechanical distribution $|\psi^*\psi|^2$. The continuity equation then gurantees that the distribution matches the quantum mechanical distribution at all future times. 

Applying the $\nabla$ operator on eqn. (\ref{3}) and remembering that $p_i = \partial_i S$, one has the second order equation of motion
\beq
\frac{dp_i}{dt} = -\partial_i \left(V + (1 - \lambda(t))Q \right).\label{9}
\eeq

This mathematical framework provides a unified basis to study continuous transitions between classical and quamtum systems via mesoscopic domains. 

\section{Equivalent Formalism Using Currents}
Using the definition of the velocity (\ref{5}) in the continuity equation (\ref{4}), it is possible to express the velocity in terms of the current density  in the form
\ben
u_i &=& \frac{d x_i}{d t} = \frac{j_i}{\rho},\label{6}\\
j_i &=& \frac{i\hbar}{2m}\left(\psi^*\partial_i \psi - \partial_i \psi^* \psi\right),\\
\rho &=& \psi^* \psi.
\een
The equation of motion is then
\beq
\frac{d p_i}{d t} = m\frac{d u_i}{d t} = m \frac{d}{d t}\left(\frac{j_i}{\rho} \right) = Q_i.\label{7} 
\eeq
Comparing with eqn. (\ref{9}), it follows that
\beq
Q_i = - \partial_i\left(V + (1 - \lambda(t)\right)Q.
\eeq

This provides a simple unified treatment for relativistic and non-relativistic particles without explicitly using the Hamilton-Jacobi method which is cumbersome in the relativistic spin-1/2 case.
Let us consider a relativistic particle described by the Dirac equation
\beq
i\hbar \frac{d\psi}{dt} = (- i\hbar \alpha_i \partial_i + \beta m_0 c^2 + V)\psi.
\eeq 
Its 4-velocity can be defined by the guidance condition
\ben
u_\mu &=& \frac{dx_\mu}{d\tau} = \frac{j_\mu}{\rho},\label{gc}\\
j_\mu &=& \bar{\psi}\gamma_\mu\psi,\\
\rho &=& \psi^\dagger \psi.
\een
The Dirac current is conserved and one has the continuity equation
\ben
\frac{\partial \rho}{\partial t} + {\bf \nabla}. {\bf j} &=& 0,\\
j_i &=& \bar{\psi}\gamma_i\psi = \rho u_i.
\een
Integration of the differential equation (\ref{gc}) for $\mu = i (i=1,2,3)$ gives a set of Bohmian trajectories of the particle corresponding to arbitrary initial positions. Now,
\beq
\frac{dp_\mu}{d\tau} = m_0 \frac{d u_\mu}{d\tau} = m_0\frac{d}{d\tau}\left(\frac{j_\mu}{\rho}\right) \equiv  Q_\mu.\label{8}
\eeq
The corresponding classical equation of motion is
\beq
\frac{dp^{cl}_\mu}{d\tau} = 0.
\eeq
Hence, the equation of motion for a mesoscopic Dirac system can be written as
\beq
\frac{dp_\mu}{d\tau} = (1 - \lambda(t))Q_\mu.
\eeq

\section{Examples}
In this section three examples will be worked out in detail to show how Bohmian trajectories smoothly pass over to classical trajectories. The motion of the particle is inextricably linked with the structure of its environment through the quantum potential $Q$. The quantum potential does not depend on the intensity of the wave, but rather on its form. It need not fall off with increasing distance. Any change in the experimental setup affects the trajectory. Therefore, the trajectories cannot be measured directly. For simplicity we put the masses of the particles as well as $\hbar$ equal to one. 

Before proceeding further, it is necessary to specify explicitly the nature of the effective coupling function
\beq
P(t) = 1 - \lambda(t).
\eeq
In order to do that, let us assume that the environment to which the quantum system is to be coupled has a random variable with a gaussian distribution. Then the cumulative distribution function (CDF) is given by
\ben
\Phi(t) &=& \frac{1}{\sqrt{2\pi\sigma^2}}\int_{-\infty}^t e^{-(x - \mu)^2/2\sigma^2}dx\\
&=& \frac{1}{2}\left[1 + {\rm erf}\left(\frac{t - \mu}{\sigma\sqrt{2}}\right)\right].
\een
This is the cumulative probability that the random variable has a value less than or equal to $t$. We will assume that $\lambda(t) = \Phi(t)$. Then $P(t)$ is the complementary CDF (CCDF) which is the cumulative probability that the random variable has a value greater than $t$. $P(t) \rightarrow 1$ as $t\rightarrow -\infty$ and $P(t) \rightarrow 0$ as $t\rightarrow\infty$. This will describe the quantum to classical transition as the system is coupled to an environment with a random variable. The reverse transition from the classical to the quantum is not physically possible because of the introduction of randomness and hence to increasing entropy. 

For simplicity of calculations we will assume that $\lambda(t)$ is a logistic function
\beq
\lambda(t) = \frac{1}{1 + e^{-b(t - t_0)}}
\eeq
where $b$ is an arbitrary parameter that can be varied and $t_0$ is the $t$-value of the sigmoid's midpoint.
Hence
\beq  
P(t) = \frac{1}{1 + e^{b(t - t_0)}}.
\eeq
A system is quantum-like for $b>0,t - t_0 <0$ and classical-like for $b>0,t - t_0 >0$, the point of inflection being $t - t_0 = 0$. 

\subsection{Double-slit}
 The quantum interference effect of matter for Young's double-slit experiment in optics is one of the famous examples. It shows how different quantum mechanics is compared to classical mechanics. Quantum particles are emitted by a source $S$, pass through two slits with the distance $\pm X$ in a barrier and arrive at a screen. To avoid complexity two identical Gaussian profiles are assumed with identical initial width $\rho$ and group velocity $u$ in $(x,t)$-space.  According to the $\rho  \left(1+\frac{i t}{\rho ^2}\right)$ term, the probability of finding the particle within a spatial interval decreases with time as the wave packet disperses. The superposition of the two Gaussian profiles leads to the interference effect. The total wave function emerging from the slits is 
\begin{equation}
\psi (x,t)=\frac{\exp  \left(-\frac{(u~t+x-X)^2}{2 \rho ^2 \left(1+\frac{i t}{\rho ^2}\right)}-i u \left(\frac{ u~t}{2}+x-X\right)\right)+\exp  \left(-\frac{(-u~t+x+X)^2}{2 \rho ^2 \left(1+\frac{i t}{\rho ^2}\right)}+i u \left(-\frac{u~t}{2}+x+X\right)\right)}{\sqrt{\rho  \left(1+\frac{i t}{\rho ^2}\right)}}.
\end{equation}
The square of the total wave function (=probability distribution) is 
\beq
\varrho (x,t)=\frac{e^{-\frac{2 \left(x^2+(-u t+X)^2\right) \rho ^2}{t^2+\rho ^4}} \left(e^{\frac{(u t+x-X)^2 \rho ^2}{t^2+\rho ^4}}+e^{\frac{(-u t+x+X)^2 \rho ^2}{t^2+\rho ^4}}+2 e^{\frac{\left(x^2+(-u t+X)^2\right) \rho ^2}{t^2+\rho ^4}} \text{Cos}\left[\frac{2 x \left(X~t+u \rho ^4\right)}{t^2+\rho
^4}\right]\right)}{\sqrt{\frac{t^2+\rho ^4}{\rho ^2}}}
\eeq
with the quantum amplitude $R$ given by $R=\sqrt{\rho (x,t)}$.
If the velocity $u_x$ is in $x$-direction, the total phase function from which $u_x$ is derived becomes
\begin{equation}
S(x,t)=\tan^{-1}\left(\frac{e^{\frac{\rho ^2 (u t+x-X)^2}{2 \left(\rho ^4+t^2\right)}} \sin \left(T_1\right)-e^{\frac{\rho ^2 (-u t+x+X)^2}{2 \left(\rho ^4+t^2\right)}} \sin \left(T_2\right)}{e^{\frac{\rho ^2 (u t+x-X)^2}{2 \left(\rho ^4+t^2\right)}} \cos \left(T_1\right)+e^{\frac{\rho ^2 (-u t+x+X)^2}{2 \left(\rho ^4+t^2\right)}} \cos \left(T_2\right)}\right)\\
\\
\end{equation}
with 
\[T_1=\frac{t (-u t+x+X)^2}{2 \left(\rho ^4+t^2\right)}-\frac{1}{2} \tan^{-1}\left(\frac{t}{\rho ^2}\right)+u \left(-\frac{ u t}{2}+x+X\right)\] and
\[T_2=-\frac{t (u t+x-X)^2}{2 \left(\rho ^4+t^2\right)}+\frac{1}{2} \tan^{-1}\left(\frac{t}{\rho ^2}\right)+u \left(\frac{u t}{2}+x-X\right).\]
To see how the Bohmian trajectories smoothly pass over to classical  trajectories, the general equation of motion (\ref{9}) is given by
\beq
\frac{d ^2 x}{dt^2} = - \partial_x\left(V + P(t)\right) Q,
\eeq
where the environment coupling function $P(t)$ is defined by (25)
with $b\in \mathbb{R}$, $t_0\in \mathbb{R}$ and with $V=0$.
 $Q$ is the quantum potential expressed by eqn. (\ref{2}), which is extremely large for the double slit.  The general equation of motion (29) can then be numerically integrated for appropriate values of the parameters $b$, $t_0$ $X$, $\rho$, $u$ and with the correct boundary conditions (see figure 1). For $b\rightarrow \infty$ and $t_0=0$ the trajectories become classical because $P$ reduces to null, and for $b\gg 1$ and $t_0=\rightarrow \infty$, the trajectories become Bohmian because $P$ tends to $1$, the quantum limit. For increasing time the amplitude of the quantum potential $Q$ decreases.
In the classical case the trajectories become straight lines because of the uniform unaccelerated motion. They cross the axis of symmetry. 
In the quantum mechanical limit the trajectories run to the local maxima of the squared wave function, which correlates with the ``plateaux'' of the quantum potential, and therefore correspond to the bright fringes of the diffraction pattern. The fate of a particle depends sensitively on its position. The quantum particle passes through slit one or slit two and never crosses the axis of symmetry. In the initial process the quantum potential is an inverse parabola associated with the Gaussian profile (see figure $4$), corresponding to the single slit Gaussian profiles. But at later times it becomes more complex and affects the motion in a considerably complicated way. The ``kinks'' in the trajectories are due to the interference coming from the part of the wave which does not carry the particle, and which is in general called an `empty wave'. In the presence of ``troughs'' the particle accelerates or decelerates. At some time steps and for special values of the parameter $b$ and $t_0$ the quantum potential blows up and accelerates the particle, which leaves the bulk of the wave packet (see for example figure $1$ with $b=0.5$ and $t_0=2$ and figure $4$). The structure of the quantum potential finally decays into a set of ``troughs'' and ``plateaux''. For further details, see the simulation in \cite{bloh1}. In figure 1 the trajectories per unit length are proportional to the probability density. The total number of trajectories is $50$.
For the Bohmian trajectories the initial particle velocities are restricted to the values $\overset{\rightharpoonup }{u}=\overset{\rightharpoonup }{\nabla } S(x_0,t=0)$.
 For the calculation the initial width $\rho$, the wavenumber $u$ and the slit distance $\pm X$ are chosen as  $\rho=\frac{5}{8}$, $u=-2$ and $X=\pm2.5$.
 In figure 2 it is seen that the initial velocity becomes $u_x(x,0)=u$ for $x<0$ and $u_x(x,0)=-u$ for $x>0$ asymptotically.
 
  \subsection{2D Harmonic Oscillator}
The time-independent wave function for a harmonic oscillator potential is a much simpler example than the double- slit experiment. A normalized, degenerate superposition of the ground state and a  first excited state with a constant relative phase shift of an uncoupled isotropic harmonic oscillator in the two-dimensional configuration space $(x,y)$ gives an entangled, stationary wave function of the form 
\begin{equation}
\psi (x,y,t)=e^{-i E_{n,m} t} \sum _{n,m} c_{n,m} \Theta_{n,m}
\end{equation}
with the the abbreviations $c_{n,m} \in \mathbb{C}$,  $\Theta_{n,m}=\phi_n(x) \phi_m(y)$ and $E_{k,j}$, where $ \phi _n (x)$, $ \phi _m (y)$ are eigenfunctions and $E_{n,m}=E_n+E_m$ are permuted eigenenergies of the corresponding one-dimensional Schr\"odinger equation.
The two-dimensional Schr\"odinger equation 
\begin{equation}
-\frac{1}{2}\left(\frac{\partial ^2}{\partial x^2}+\frac{\partial ^2}{\partial y^2}\right)\psi +V(x,y)\psi =i\frac{\partial }{\partial t}\psi
\end{equation}
leads to the degenerate, entangled wave function $\Psi$ for the  harmonic potential $V(x,y)=\frac{1}{2} k_0 \left(x^2+y^2\right)$.
In general a stationary wave function yields a stationary velocity field. For a degenerate superposition of two eigenstates the total phase function becomes independent of the variable $x$ and therefore the velocity equals null. To get an autonomous velocity with a non-trivial motion a constant phase shift is introduced. Phase shift techniques have been widely used in modern phase measurement instruments. It is called \textit{Phase Shift Interferometry} (PSI). The basic idea of the PSI technique is to vary the phase between two beams in some manner to get a phase difference. For instance, the Mach-Zehnder interferometer is a device used to determine the relative phase shift between two beams derived by splitting the beam from one single source. The interferometer has been applied to determine phase shifts between the two beams caused by a change in length of one of the arms. In our special case the phase timeshift is introduced by the factor $\alpha$ with $\alpha = t_1$, which leads to stationary wave function for the harmonic oscillator:

\begin{equation}
\psi (x,y,t)=e^{i \alpha}\Theta _{0,1}\text{ + }\Theta _{1,0}
\end{equation}
with the  angular frequency $\omega =\sqrt{k_0}$ and with the eigenfunctions:
\begin{equation}
\Theta _{m,n}=\frac{\omega ^2 H_m\left(\omega ^2 y\right) H_n\left(\omega ^2 x\right) \exp \left(-\frac{1}{2} \omega  \left(x^2+y^2\right)- i~ \omega t~ ( m+ n+1)\right)}{\sqrt{2 \pi } \sqrt{2^m m!} \sqrt{2^n n!}}.
\end{equation}
For the trajectories in figure 3, $\alpha=\frac{\pi}{2}$, $k_0=1$ and $e^{\frac{i \pi}{2}}=i$. Therefore the wave function becomes
\begin{equation}
\psi (x,y,t)=\text{ = }\frac{1}{\sqrt{\pi }}
(x+\text{iy}) e^{-\frac{1}{2}  \left(4 i t+x^2+y^2\right)}.
\end{equation}

For the general case (28) the square of the total wave function (=probability distribution) becomes time-indepedent:
 \begin{equation}
 \varrho (x,y)=\frac{1}{\pi }e^{-\omega  \left(x^2+y^2\right)}\omega ^2 \left(x^2+y^2+2~ x~ y~ \cos(\alpha)\right)
 \end{equation}
 The phase function $S$ from this total wave function is
 \begin{equation}
 S(x,y,t)=-\tan^{-1} \left( \frac{x~ (\sin  (2 ~ \omega t))+y~ (\sin  (2 ~\omega t -\alpha ))}{x~ (\cos  (2 ~ \omega t))+y~ (\cos  (2 ~ \omega t-\alpha ))} \right).
 \end{equation}
From the gradient of the phase function $S$ we get the corresponding autonomous differential equation system for the velocity field $\overset{\rightharpoonup }{u}$ in the $x$ and $y$ -directions:
   \begin{eqnarray}
  u_x(x,y)=-\frac{y \sin (\alpha)}{x^2+2~ x~ y~ \cos ( \alpha   )+y^2} \nonumber\\
  u_y(x,y)=~\frac{x \sin (\alpha)}{x^2+2~ x~ y~ \cos ( \alpha   )+y^2}.
  \end{eqnarray}
The velocity vector is independent of the parameter $\omega$ and can be integrated numerically with respect to time to yield the motion in the ($x$,$y$) configuration space.
The origin of the motion lies in the relative phase of the total wavefunction, which has no classical analogue in particle mechanics.

With the quantum amplitude (=absolute value)$R=\sqrt{\rho (x,y)}$, 
the quantum potential is:
\begin{equation}
Q(x,y)=\frac{-T_1 ~ \left(x^2+y^2\right)-4~ T_2~ x~ \omega ~ y \cos (\alpha ) \left(x^2+y^2\right)+T_3 \cos ^2(\alpha )}{2 \left(x^2+2 x y \cos (\alpha )+y^2\right)^2}
\end{equation}
with $T_1 =\omega ^2~ \left(x^2+y^2\right)^2-4~ \omega  \left(x^2+y^2\right)+1$, $T_2=\omega ~ \left(x^2+y^2\right)-4$ and \\ $T_3 =x^2 \left(-4~ \omega ^2 y^2 \left(x^2+y^2\right)+16 \omega  y^2+1\right)+y^2$.
For the trajectories in figure 3 the quantum potential becomes with $\alpha=\frac{\pi}{2}$ and $\omega=1$ (see figure $5$):
\begin{equation}
Q(x,y)=-\frac{ \left(x^2+y^2\right){}^2-4 ~ \left(x^2+y^2\right)+1}{2 \left(x^2+y^2\right)}.
\end{equation}
For $b\approx 1$ (as well as $b \gg 1$) and $t_0\rightarrow \infty$  the Bohmian trajectory forms a closed circle when the trajectory is projected in the $xy$ plane (see figure $3$) and for $b=46$ and $t_0=0$ a closed ellipse is obtained (the classical limit). For the classical case the equations of motion can be easily integrated to yield $x(t)=x_m \cos( \omega~ t-\phi_x)$ and $y(t)=y_m \cos( \omega~ t-\phi_y)$, where $x_m$, $y_m$, $\phi_x$ and $\phi_y$ are real valued constants. The direction and form (circle or ellipse) of the orbits depend on the phase difference $\phi_x-\phi_y$. Increasing the time for the mesoscopic case (for example $b=16$ and $t_0=10$ )the trajectories evolve after some time steps into the classical motion(ellipse). The orbit is substantially influenced by the initial starting point (small black points in figure $3$) of the particles. In some mesoscopic cases the orbit is not closed for certain time intervals (see figure $3$: $b=0.5$ and $t_0=4$) but no exponential separation of neighboring orbits appears(chaotic motion). Chaotic motion arises because of the nodal point (singularity) of the wave function. At the nodal point, the quantum potential becomes very negative or approaches negative infinity, which keeps the particles from entering the nodal region (for further details of the Bohmian trajectories, see for example \cite{bloh2}).
\subsection{The Hydrogen-like Atom}
The hydrogen atom contains a single positively charged proton and a single negatively charged electron bound to the nucleus by the Coulomb force.
The solution of the Schrodinger equation in spherical polar coordinates for the hydrogen atom is accessed by separating the variables. The wave function is represented by the product of functions $\psi(\rho,\theta ,\phi ,t)=R_{n,l}(r) Y_{lm}(\phi,\theta)e^{i E_n t}$. The solutions can exist only when constants which arise in the solution, are restricted to integer values: The principal quantum number n, the angular momentum quantum number or azimunthal quantum number l and magnetic quantum number m. The potential, $V$ between two charges is described by  Coulomb's law:
\begin{equation}
V=-\frac{e^2 Z}{4 \pi  r \epsilon _0}
\end{equation}
where $Ze$ is the charge of the nucleus (Z=1 being the hydrogen case, Z=2 helium, etc.), the $e$ is the elementary charge of the single electron, $\epsilon_0$ is the permittivity of vacuum. With the system consisting of two masses,  the reduced mass is defined by:
\begin{equation}
 \mu =\frac{m M}{m+M}
 \end{equation}
 where $M$ is the mass of the nucleus and $m$ the mass of the electron. If the nucleus is much more massive than the electron the reduced mass becomes $\mu \approx m$.
 The  Hamiltonian for the hydrogen atom is:
 \begin{equation}
  H=-\frac{\hbar ^2}{2 \mu }\nabla ^2-\frac{e^2 Z}{4 \pi  r \epsilon _0}.
  \end{equation}
  For simplicity we put the reduced masses $\mu$ of the particles, $\hbar$ as well as $\frac{e^2 Z}{4 \pi  \epsilon _0}$ equal to one.
  Since the potential is spherically symmetric, the Schrodinger equation can be solved analytically in spherical polar coordinates. This leads to the time-dependent wafefunction with the associated Laguerre polynomials $L_{-l+n-1}^{2 l+1}(\rho)$ and the Laplace's spherical harmonics $Y_l^m(\theta ,\phi)$:

  \begin{equation}
 \psi (\rho ,\theta ,\phi ,t)=  2\rho ^l e^{-\frac{\rho }{2}-i E_n t}\sqrt{\frac{(-l+n-1)!}{n^4 (l+n)!}} L_{-l+n-1}^{2 l+1}(\rho)Y_l^m(\theta ,\phi )
   \end{equation}
   with the energy eigenstates $E_n$, diameter $\rho =2 r$ where the radius $r$ is the distance between the nucleus and the electron.
 The energy levels $E_n$ of the hydrogen atom depends on the principal quantum number $n$:
  \begin{equation}
  E_n=-\frac{Z^2 e^4 \mu}{2 (4 \pi)^2 \epsilon_0^2 \hbar^2 n^2}
  \end{equation}
 which is in our case simplified to $E_n=-\frac{1}{2 n^2}$. A stationary state $\psi (\rho ,\theta ,\phi ,t)$ corresponding to the energy state $E_n$ leads to the phase function $S$
 \begin{equation}
   S(\rho ,\theta ,\phi ,t)=E_n t+m \hbar \phi.
   \end{equation}  
For each $t$ and $m \neq 0$ the electron rotates about the z-Axis \cite{Holland}. To give an example of the smooth transition a stationary state with the quantum numbers $n=2$, $l=1$ and $m=1$ and cartesian coordinates is applied.
In such a case the sperical variables are transformed via
$r=\sqrt{x^2+y^2+z^2}$, $\theta=\tan^{-1}\left( \frac{\sqrt{x^2+y^2}}{z } \right)$ and $\phi=\tan^{-1}\left(\frac{y}{x}\right)$ which lead to the simple wavefunction $\psi$
\begin{equation}
\psi=-\frac{(x+i y) e^{-\frac{1}{2} \sqrt{x^2+y^2+z^2}+\frac{i t}{8}}}{8 \sqrt{\pi }}.
\end{equation}
From the wavefunction $\psi$ it follows for the phase function $S$ 
\begin{equation}
S(x,y,z,t)= \tan ^{-1}\left(\frac{x \sin \left(\frac{t}{8}\right)+y \cos \left(\frac{t}{8}\right)}{x \cos \left(\frac{t}{8}\right)-y \sin \left(\frac{t}{8}\right)}\right)
\end{equation}
and therefore for the $(x,y,z)$ - components of the velocity
\begin{eqnarray}
v_x&=&-\frac{y}{x^2+y^2}\\ \nonumber
v_y&=&\frac{x}{x^2+y^2}\\ \nonumber
v_z&=&0.
\end{eqnarray}
The particle rotates with a constant speed $v_x^2+v_y^2$ which is proportional to $\frac{1}{r^2}$. That means if the particle is nearer to the nucleus it moves faster.
From the quantum amplitude $R$ with $R=\frac{\sqrt{x^2+y^2} e^{-\frac{1}{2} \sqrt{x^2+y^2+z^2}}}{8 \sqrt{\pi }}$ the quantum potential $Q$ becomes
\begin{equation}
Q(x,y,z)=-\frac{x^2 \left(1-\frac{8}{\sqrt{x^2+y^2+z^2}}\right)+y^2 \left(1-\frac{8}{\sqrt{x^2+y^2+z^2}}\right)+4}{8 \left(x^2+y^2\right)}.
\end{equation}
With the classical potential $V=-\frac{1}{\sqrt{x^2+y^2+z^2}}$ and with the logistic function $\lambda (t)$ the smooth transition could be calculated. 
In figure $6$ you see the classical motion for 3 orbits with different starting positions and with the initial velocities taken from wave density and the gradient of the phase function for the quantum case. In this case the orbits are closed ellipses in the $(x,y,z)$ - space. Ignoring Pauli's exclusion principle that states that two or more identical electrons cannot occupy the same quantum state within a quantum system simultaneously you see in figure $7$ the "quantum motion," as represented by 3 Bohm trajectories with the wave density contours and the quantum potential contours. The smooth transition from the quantum to the classical case is shown in figure 8 where it is clearly seen that the quantum motion, which are circles in the $(x,y)$ - space perpendicular to the z-axis passes over to the ellipses in the $(x,y,z)$ - space.

 \subsection{Remark on the numerical methods}
 The trajectories are calculated with different methods by \textit{Mathematica}'s built in function \textbf{NDSolve} and compared with different numerical methods, offered by the method option in NDSolve, e. g. Runge$-$Kutta methods or the midpoint method. The accuracy for the calculations is 7 digits, which specifies how many effective digits of accuracy should be sought in the final result and with the maximum number of 50000 steps to generate a result.
 \section{Acknowledgement}
PG acknowledges financial support from the National Academy of Sciences, India.     

\pagebreak
\begin{figure}
\begin{center}
{\includegraphics[scale=0.45]{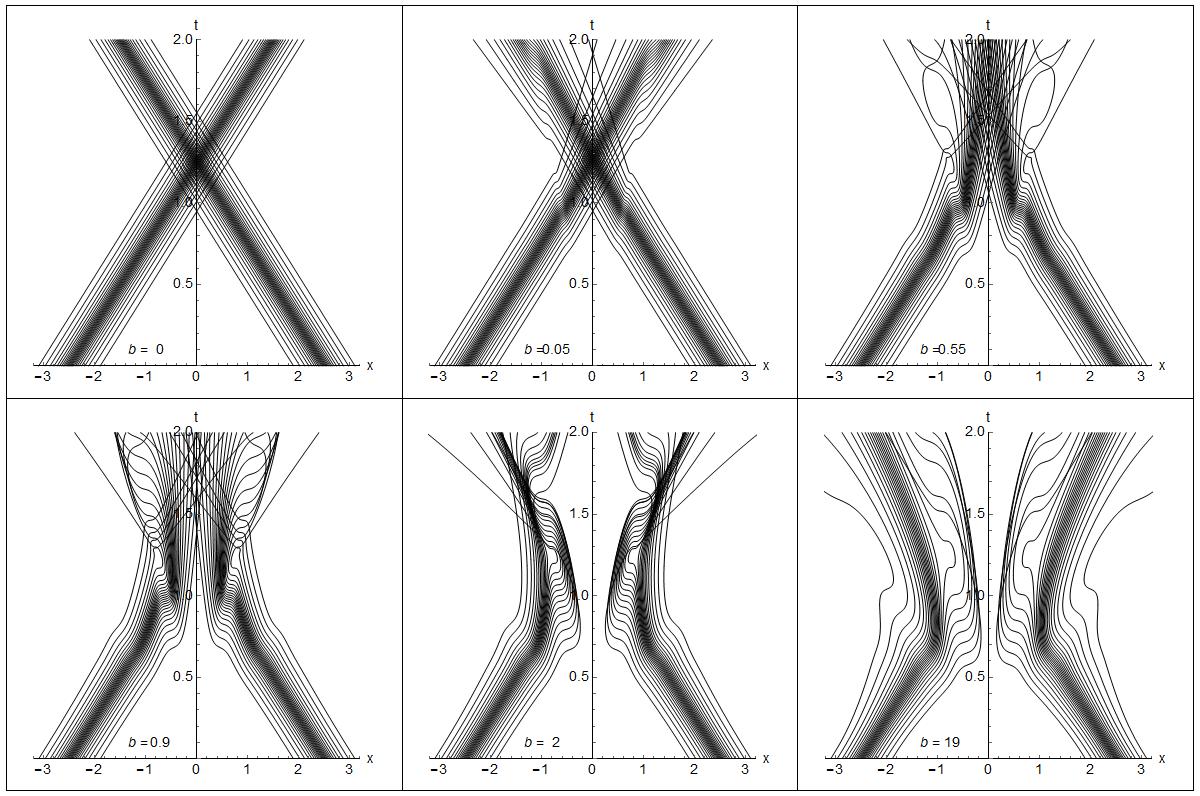}}
\caption{\label{Figure 1}{\footnotesize Six snapshots of the double-slit arrangement, showing how the Bohmian trajectories ($b=15$, $t_0=2$) continuously become classical trajectories via mesoscopic states as $t_0$ decreases to 0 and $b$ increases to 40. The parameters are: time $0 \le t \le 2.0 $, initial density $\rho=\frac{5}{8}$, wavenumber $u=-2$ and slit distance $X=\pm2.5$}}
\end{center}
\end{figure}
\pagebreak

 \begin{figure}
 \begin{center}
  {\includegraphics[scale=0.7]{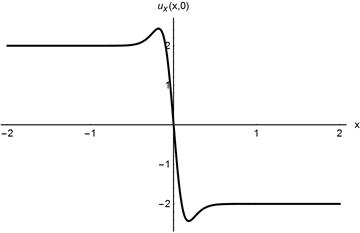}}
\caption{\label{Figure 2}{\footnotesize Initial velocity/wavenumber}}
\end{center}
\end{figure}
\pagebreak

\begin{figure}
 \begin{center}
  {\includegraphics[scale=0.6]{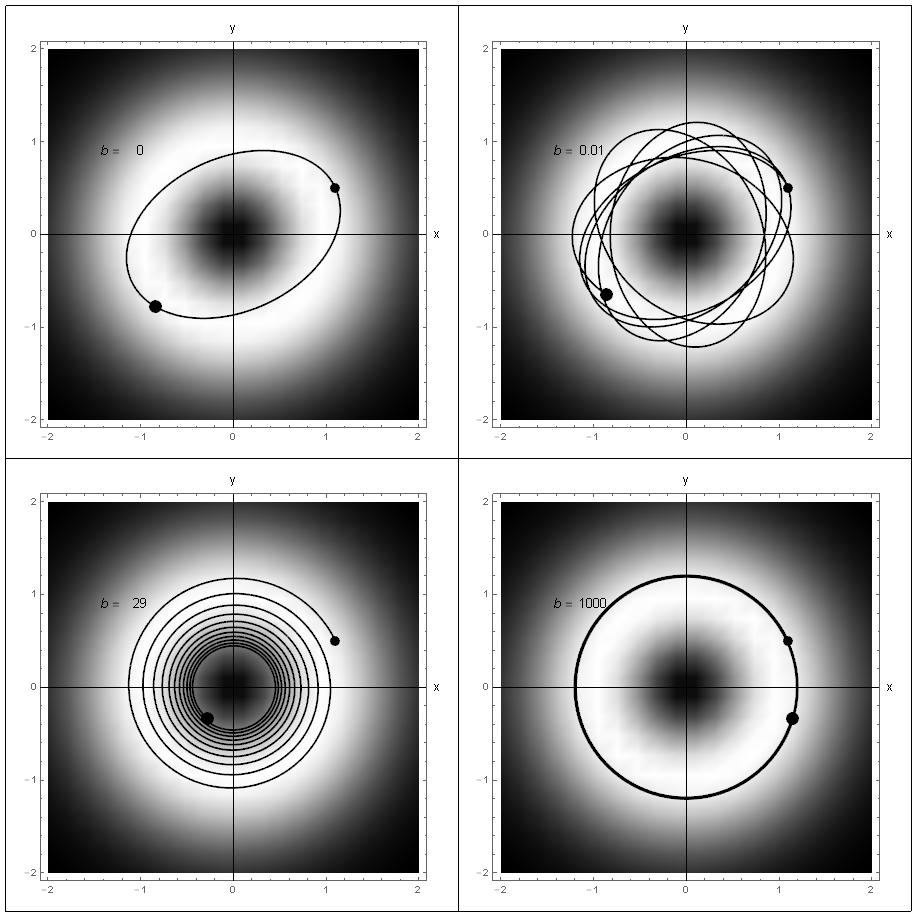}}
\caption{\label{Figure 3}{\footnotesize Four snapshots of the 2D harmonic oscillator trajectories showing the transition from Bohmian ($b=1$, $t_0=200$) to classical trajectories ($b=37$, $t_0=0$) via mesoscopic states as $t_0$ decreases to 0. The parameter are: time $0 \le t \le 35 $, constant $k_0=1$, initial point: small black point}}
\end{center}
\end{figure}

\begin{figure}
 \begin{center}
  {\includegraphics[scale=1.2]{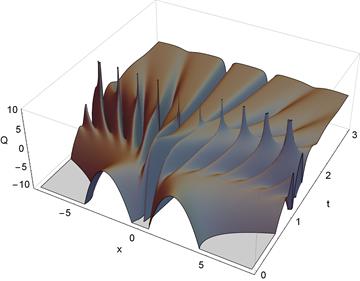}}
\caption{\label{Figure 4}{\footnotesize The quantum potential for the two-slit experiment. The parameters are: $\rho=\frac{5}{8}$, $u=-2$, $X=\pm2.5$ and time $0 \le t \le 2.5 $ }}
\end{center}
\end{figure}

\begin{figure}
 \begin{center}
  {\includegraphics[scale=1.2]{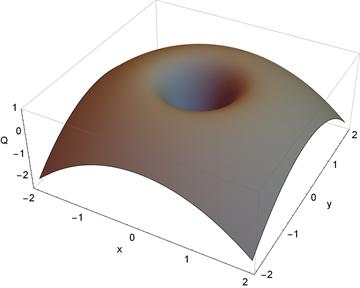}}
\caption{\label{Figure 5}{\footnotesize The quantum potential for the harmonic oscillator. The parameters are: $k_0=1$ and $\alpha=\frac{\pi}{2}$  }}
\end{center}
\end{figure}

\begin{figure}
 \begin{center}
  {\includegraphics[scale=0.6]{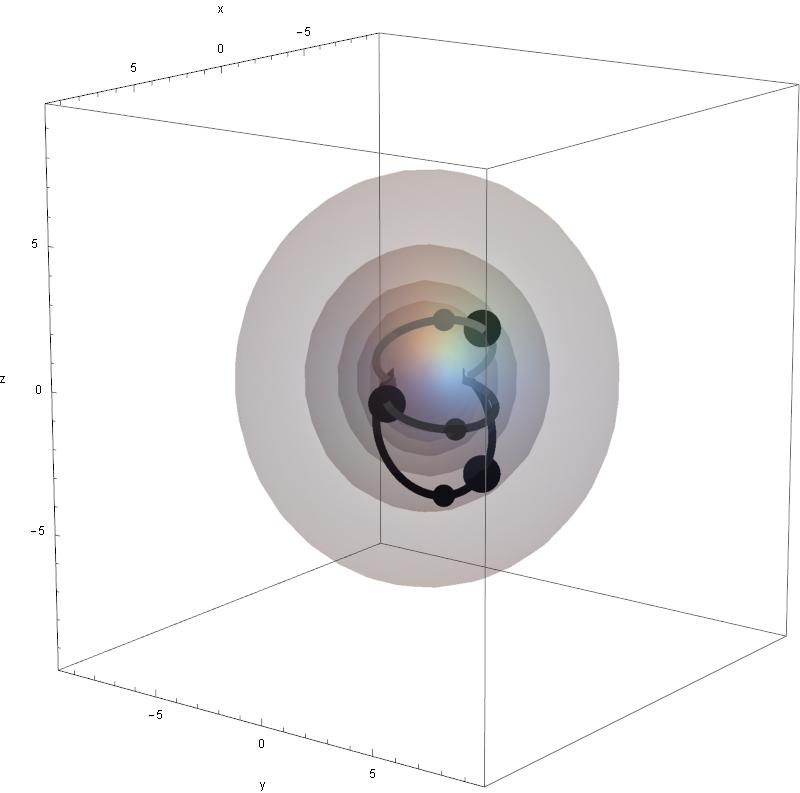}}
\caption{\label{Figure 6}{\footnotesize The Coulomb potential and 3 classical orbits with initital position and velocities taken from wave density and the gradient of the phase function for the quantum case. The parameters are: time $0 \le t \le 2500 $ and initial points: please, see small points}}
\end{center}
\end{figure}

\begin{figure}
 \begin{center}
  {\includegraphics[scale=0.6]{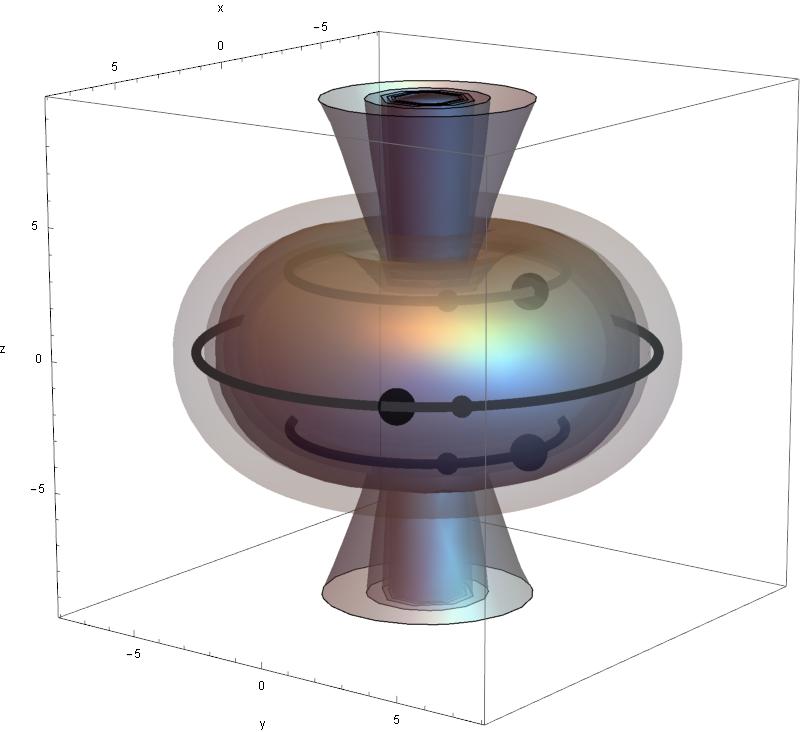}}
\caption{\label{Figure 7}{\footnotesize The wave density, the quantum potential and 3 Bohmian trajectories with  different initital positions and velocities. The parameters are: time $0 \le t \le 2500 $ and initial points: please, see small points}}
\end{center}
\end{figure}

\begin{figure}
 \begin{center}
  {\includegraphics[scale=0.5]{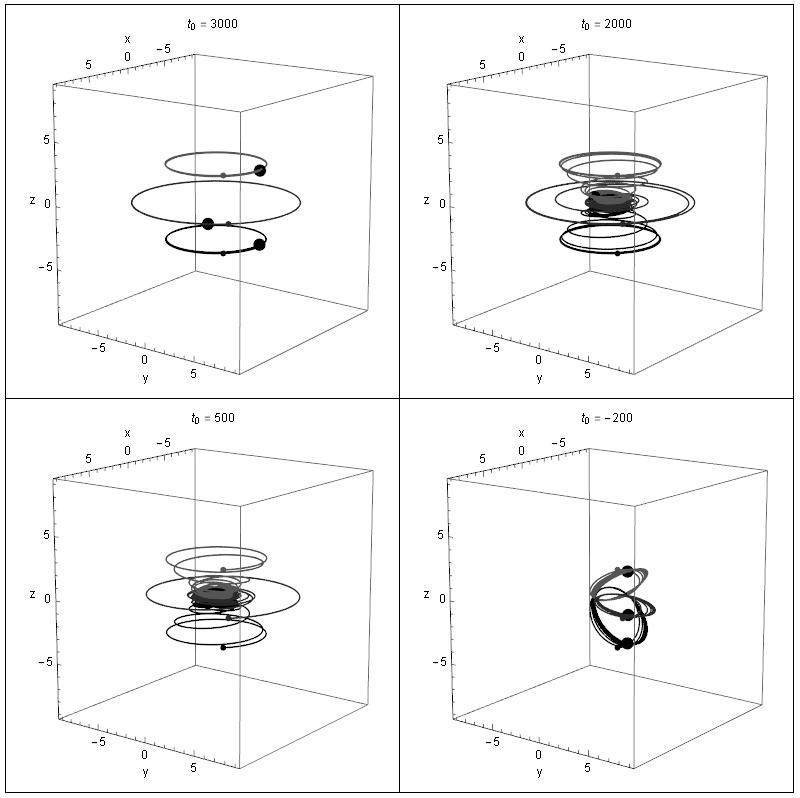}}
\caption{\label{Figure 8}{\footnotesize Smooth transition for different $t_0$. The parameters are: time $0 \le t \le 2500 $, $b=0,014$, end points: please, see big points and initial points: please, see small points}}
\end{center}
\end{figure}

\end{document}